\documentclass[graybox]{svmult}

\usepackage{comment}
\usepackage{type1cm}        % activate if the above 3 fonts are
\usepackage{makeidx}         % allows index generation
\usepackage{graphicx}        % standard LaTeX graphics tool
\usepackage{multicol}        % used for the two-column index
\usepackage[bottom]{footmisc}% places footnotes at page bottom

\usepackage[T1]{fontenc}      % encoding for proper glyph selection
\usepackage[utf8]{inputenc}   % if you compile with pdfLaTeX
\usepackage{newtxtext}       % 
\usepackage[varvw]{newtxmath}       % selects Times Roman as basic font
\usepackage{array}     % m{<width>} for vertical centring
\usepackage{enumitem}  % tidy bullet lists
\usepackage{svg}
\setsvg{inkscapelatex=false}  % Prevent errors from LaTeX text in SVG

\makeindex             % used for the subject index
\begin{document}

%\title*{AI-Driven Post-Quantum Cryptography for Cyber-Resilient Communication in Transportation Cyber-Physical Systems}
\title*{AI-Driven Post-Quantum Cryptography for Cyber-Resilient V2X Communication in Transportation Cyber-Physical Systems}

\titlerunning{AI-Driven PQC for Cyber-Resilient V2X Communication in TCPS}
\author{Akid Abrar,\orcidID{0009-0007-5880-9389} \\ 
Sagar Dasgupta, Ph.D.\orcidID{0000-0001-8491-662X}\\
Mizanur Rahman, Ph.D.\orcidID{0000-0003-1128-753X}\\
Ahmad Alsharif, Ph.D.\orcidID{0000-0003-1060-1953}}
\institute{Akid Abrar \at Ph.D. Student in Transportation Systems Engineering, Department of Civil, Construction, and Environmental Engineering, the University of Alabama, Tuscaloosa, AL, USA. \email{aabrar@crimson.ua.edu} 
\and Sagar Dasgupta, Ph.D. \at Postdoctoral fellow at Connected and Automated Mobility Laboratory, the University of Alabama, Tuscaloosa, AL, USA. \email{sdasgupta@ua.edu}
\and Mizanur Rahman, Ph.D. \at Assistant Professor in Transportation Systems Engineering, Department of Civil, Construction and Environmental Engineering, the University of Alabama, Tuscaloosa, AL, USA. \email{mizan.rahman@ua.edu}
\and Ahmad Alsharif, Ph.D. \at Assistant Professor, Department of Computer Science, the University of Alabama, Tuscaloosa, AL, USA. \email{ahmad.alsharif@ua.edu}}
\maketitle

\abstract*{Transportation Cyber-Physical Systems (TCPS) integrate physical elements, such as transportation infrastructure and vehicles, with cyber elements via advanced communication technologies 
% \textcolor{blue}{\textbf{Avoid using vague terms like “advanced communication technologies” — be specific about the technologies being discussed.}}
, allowing them to interact seamlessly. This integration enhances the efficiency, safety, and sustainability of transportation systems. TCPS rely heavily on cryptographic security to protect sensitive information transmitted between vehicles, transportation infrastructure, and other entities within the transportation ecosystem, ensuring data integrity, confidentiality, and authenticity. Traditional cryptographic methods have been employed to secure TCPS communications, but the advent of quantum computing presents a significant threat to these existing security measures. Therefore, integrating Post-Quantum Cryptography (PQC) into TCPS is essential to maintain secure and resilient communications. While PQC offers a promising approach to developing cryptographic algorithms resistant to quantum attacks, artificial intelligence (AI) can enhance PQC by optimizing algorithm selection, resource allocation, and adapting to evolving threats in real-time. AI-driven PQC approaches can improve the efficiency and effectiveness of PQC implementations, ensuring robust security without compromising system performance. This chapter introduces TCPS communication protocols, discusses the vulnerabilities of corresponding communications to cyber-attacks, and explores the limitations of existing cryptographic methods in the quantum era. By examining how AI can strengthen PQC solutions, the chapter presents cyber-resilient communication strategies for TCPS.}

\abstract{Transportation Cyber-Physical Systems (TCPS) integrate physical elements, such as transportation infrastructure and vehicles, with cyber elements via advanced communication technologies 
% \textcolor{blue}{\textbf{Avoid using vague terms like “advanced communication technologies” — be specific about the technologies being discussed.}}
, allowing them to interact seamlessly. This integration enhances the efficiency, safety, and sustainability of transportation systems. TCPS rely heavily on cryptographic security to protect sensitive information transmitted between vehicles, transportation infrastructure, and other entities within the transportation ecosystem, ensuring data integrity, confidentiality, and authenticity. Traditional cryptographic methods have been employed to secure TCPS communications, but the advent of quantum computing presents a significant threat to these existing security measures. Therefore, integrating Post-Quantum Cryptography (PQC) into TCPS is essential to maintain secure and resilient communications. While PQC offers a promising approach to developing cryptographic algorithms resistant to quantum attacks, artificial intelligence (AI) can enhance PQC by optimizing algorithm selection, resource allocation, and adapting to evolving threats in real-time. AI-driven PQC approaches can improve the efficiency and effectiveness of PQC implementations, ensuring robust security without compromising system performance. This chapter introduces TCPS communication protocols, discusses the vulnerabilities of corresponding communications to cyber-attacks, and explores the limitations of existing cryptographic methods in the quantum era. By examining how AI can strengthen PQC solutions, the chapter presents cyber-resilient communication strategies for TCPS.}

%For example, Shor’s algorithm can compromise widely used cryptographic schemes, including Rivest–Shamir–Adleman (RSA), Diffie-Hellman (DH), Digital Signature Algorithm (DSA), and Elliptic Curve Cryptography (ECC). Additionally, Grover’s algorithm accelerates brute-force attacks on symmetric encryption methods, such as Advanced Encryption Standard (AES), making TCPS vulnerable to data breaches.

%The goal is to inform researchers, practitioners, and policymakers about the necessity of adopting AI-driven PQC strategies to protect TCPS against future cyber threats, thereby ensuring system reliability, safeguarding user data, and maintaining public trust in an increasingly connected world.

\section{Introduction}
\label{sec:1}

Transportation Cyber-Physical Systems (TCPS) represent a transformative paradigm in modern mobility infrastructure by tightly integrating physical components, such as vehicles, traffic signals, roadside sensors, and pedestrian devices, with cyber elements including edge computing, cloud platforms, and high-speed wireless communication networks ~\cite{Moller2016}. These systems enable real-time perception, decision making, and control between distributed agents in the transportation ecosystem. As urbanization accelerates and connected and automated vehicles (CAVs) proliferate, TCPS are becoming the backbone of intelligent transportation systems (ITS), promising enhanced safety, efficiency, sustainability, and user experience.\\

At the heart of TCPS lies its communication architecture, which enables seamless information exchange among vehicles, infrastructure, pedestrians, and back-end services. This architecture is collectively referred to as Vehicle-to-Everything (V2X) communication. Within V2X, several distinct yet interconnected modalities operate, including Vehicle-to-Vehicle (V2V), Vehicle-to-Infrastructure (V2I), Vehicle-to-Network (V2N) and Vehicle-to-Pedestrian (V2P) links, and it supports various safety, mobility and energy-efficient applications. For example, V2V supports cooperative maneuvers and collision avoidance by enabling nearby vehicles to share real-time motion data. V2I facilitates bidirectional communication between vehicles and roadside infrastructure (e.g., traffic signals and Roadside Units (RSUs)). On the other hand, V2N provides connectivity to cloud services and traffic management centers for functions such as over-the-air updates and dynamic routing. V2P enhances safety for vulnerable road users through interactions with smartphones or wearable devices. These V2X modalities collectively form the backbone of connected mobility within TCPS.\\

Together, these V2X modalities form a multilayered data-sharing ecosystem that supports cooperative perception, distributed decision making, and adaptive infrastructure response. The practical benefits of this system are compelling. According to the National Highway Traffic Safety Administration (NHTSA), the V2V and V2I communication systems, key elements of TCPS, could prevent or mitigate up to 80\% of crashes not involving impaired drivers~\cite{NHTSA80percent}. Independent modeling on Australian crash data suggests injury crashes could fall by 43–55\% and fatal crashes by 31–37\% once fleet penetration reaches full scale~\cite{Doecke2021} and lead to billions of dollars in annual savings through reduced traffic congestion, emissions, and fuel consumption.\\

As TCPS become increasingly reliant on continuous wireless data exchange, they also face growing exposure to cyber threats~\cite{Twardokus2024}. Safety-critical messages now carry sensitive information, such as precise geolocation, trajectory data, authentication credentials, digital payments, and even firmware updates. Ensuring the confidentiality, authenticity, and integrity of these messages is essential to prevent malicious actors from injecting false information, impersonating trusted entities, or disrupting traffic flow through jamming or spoofing attacks.\\

Currently, TCPS security relies on classical cryptographic algorithms, such as Rivest–Shamir–Adleman (RSA)~\cite{RSA}, Diffie–Hellman~\cite{Diffie}, and Elliptic Curve Cryptography (ECC)~\cite{ECDSA} for key exchange and authentication; AES~\cite{AES2001} and ChaCha20~\cite{ChaCha20} for symmetric encryption; and SHA-2/SHA-3~\cite{Zimmermann1980,SHA} for integrity protection. These cryptographic primitives have formed the foundation of secure communication protocols in vehicular networks for over a decade. However, with the rapid progress in quantum computing, their long-term viability is in question. Quantum algorithms, particularly Shor’s algorithm and Grover’s algorithm, threaten to undermine the core mathematical assumptions behind classical cryptography—making it possible for adversaries to forge signatures, decrypt secure messages, or break authentication mechanisms that protect safety-critical systems.\\

In response to these emerging cyber risks, the cryptographic community has developed a new generation of algorithms under the umbrella of Post-Quantum Cryptography (PQC).In July 2022, the U.S. National Institute of Standards and Technology (NIST) announced the first four post-quantum algorithms it will standardize: CRYSTALS-Kyber for  Key Encapsulation Mechanism (KEM), and the three digital-signature schemes, i.e., CRYSTALS-Dilithium, Falcon, and SPHINCS+. Draft Federal Information Processing Standards (FIPS 203–205) for Kyber, Dilithium and SPHINCS+ were published in 2023 and finalized in August 2024. NIST added another KEM, called HQC, as a backup on March 2025, launching work on a separate FIPS that is expected to be completed by 2027~\cite{NIST2024FIPS,HQC}. Early experiments demonstrate that these algorithms perform well on high-speed fiber links between traffic signal controllers; however, their larger keys and more complex mathematical operations can push low-power on-board units beyond the 100 ms minimum delay (computing and/or communication delay) requirements that safety messages must meet~\cite{Lonc2023}. To bridge that gap, researchers are turning to \emph{AI-driven PQC}: AI agents that monitor communication channel load, Central Processing Unit (CPU) usage, and threat level and then choose the lightest cryptographic scheme for the system security on the fly, compress keys, or off-load computationally intensive mathematical operations to roadside processing units~\cite{Ryan2025,Premakumari2025}.\\

This chapter explores the convergence of AI and PQC in the context of TCPS. Section~\ref{sec:2} provides a detailed overview of TCPS architecture and its communication vulnerabilities. Section~\ref{sec:3} examines classical cryptographic primitives and their susceptibility to quantum threats. Section~\ref{sec:4} presents the current landscape of post-quantum cryptographic algorithms and their trade-offs. Section~\ref{sec:5} introduces AI-driven strategies to enable context-aware and adaptive deployment of PQC in TCPS. Section~\ref{sec:6} outlines key research directions toward lightweight cryptographic designs, AI-augmented security orchestration, and scalable deployment frameworks. Finally, Section~\ref{sec:7} concludes the chapter with a summary of key findings.

\section{TCPS and TCPS Communication Vulnerabilities}
\label{sec:2}

A TCPS consists of diverse components, including vehicles with onboard sensors and computing units, roadside transportation infrastructure (e.g., traffic signals, sensors and roadside data infrastructure), communication mediums and networks (e.g., cellular V2X (C-V2X) and vehicular ad-hoc networks), and back-end services (e.g., traffic management centers and cloud services). Vehicles communicate with other vehicles (V2V), with infrastructure (V2I), and with cloud services or control centers (V2N)—collectively referred to as V2X communication~\cite{Moller2016}. An overall TCPS architecture typically integrates these elements so that data flows efficiently and securely among all parties~\cite{Moller2016}. Secure credential management systems (e.g., the IEEE 1609.2 WAVE/SCMS in the US) are deployed to manage digital certificates for vehicles, ensuring trust across the system~\cite{USDOT2019, IEEE1609_2022}.

\subsection{TCPS Architecture }
\label{subsec:TCPSarch}

\begin{figure}[ht]
    \centering
    \includegraphics[height=0.8\textheight]{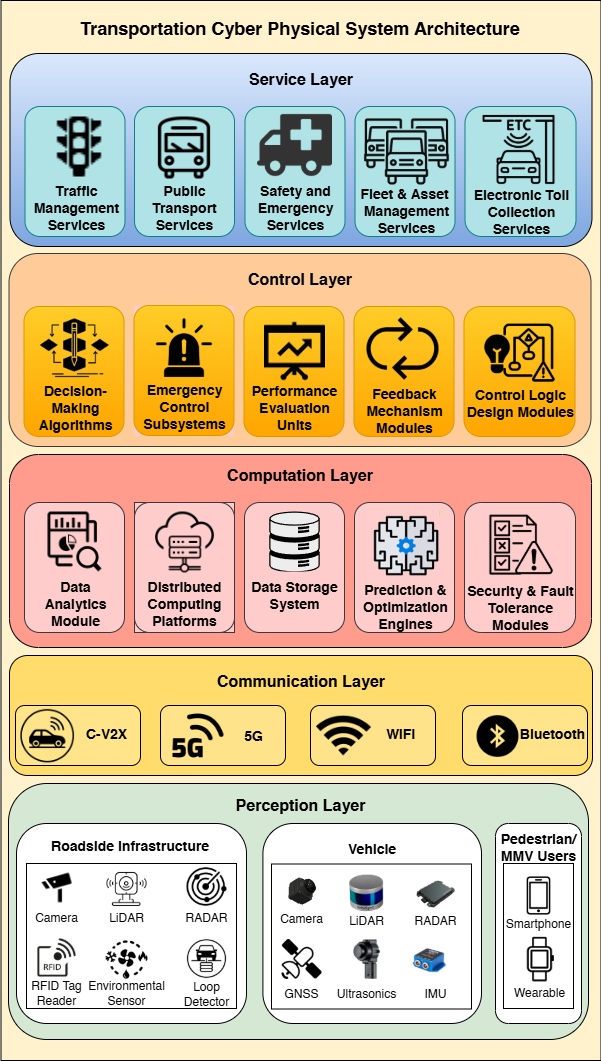}
    \caption{TCPS Architecture}
    \label{fig:TCPS-Architecture}
\end{figure}

The architecture of TCPS are defined through a layered integration of sensing, communication, computation, control, and service functionalities ~\cite{6391303}. Each layer is logically distinct yet functionally interlinked, forming a hierarchical structure that facilitates the seamless information and control signal exchange between cyber components (e.g., algorithms, software agents, and data processing platforms) and physical components (e.g., vehicles, infrastructure, mobile users) in real-time. This architecture enables system-wide situational awareness, adaptive decision-making, and responsive actuation, and is applicable to a wide range of deployment environments including urban, suburban, and rural areas (see Figure ~\ref{fig:TCPS-Architecture}). 

The Perception Layer represents the system’s primary interface with the physical transportation environment. It includes a diverse array of sensors deployed across vehicles, roadside infrastructure, and vulnerable road users, such as pedestrians and micro-mobility device operators. Roadside infrastructure typically integrates fixed-position sensors, such as camera, radar, LiDAR, RFID tag readers, inductive loop detectors, and environmental monitoring systems. These sensors monitor traffic flow, vehicle presence, occupancy rates, road surface conditions, weather parameters, and visibility. Within vehicles, on-board perception systems consist of GNSS receivers for global positioning, inertial measurement units (IMUs) for acceleration and orientation tracking, ultrasonic sensors for proximity detection, and multi-modal systems integrating LiDAR, radar, and vision-based technologies for 360-degree situational awareness. Pedestrian and Micro-Mobility Vehicle (MMV) sensing is enabled via personal smartphones and wearable devices that are capable of participating in V2P communication protocols. The data collected in this layer is typically timestamped and geo-referenced and forms the raw input for all downstream processing layers.\\

The Communication Layer ensures that data collected by the perception layer, along with messages generated by upper layers, can be transmitted reliably and with low latency across the TCPS ecosystem. This layer supports a range of wireless technologies, including Cellular Vehicle-to-Everything (C-V2X), 5G New Radio (5G NR), Wi-Fi, and Bluetooth Low Energy (BLE). The C-V2X communication stack includes two modes: the PC5 interface, which supports direct communication between nearby entities (e.g., V2V, and V2I, and the Uu interface, which enables long-range, network-based communication with cellular infrastructure (e.g., Vehicle-to-Network or V2N). Following the 2020 reallocation of the 5.9 GHz band by the Federal Communications Commission (FCC), Dedicated Short-Range Communications (DSRC) based on IEEE 802.11p has been deprecated in the United States, and new spectrum allocations have been prioritized for C-V2X operations. Key communication nodes in this layer include on-board units (OBUs) embedded in vehicles, RSUs positioned along the roadway, edge gateways, and cellular base stations. These nodes facilitate the exchange of safety-critical and operational data across V2V, V2I, V2N, and V2P domains. To maintain secure and trustworthy communication, the layer incorporates Public Key Infrastructure (PKI), digital certificate management, and cryptographic authentication and encryption mechanisms. These security measures ensure that all transmitted data is verifiable, tamper-resistant, and available in a timely manner, which is essential for enabling cooperative safety and control functionalities in connected transportation systems.\\

The Computation Layer processes, filters, fuses, and analyzes the high-volume, heterogeneous data streams originating from the perception layer. This layer spans three tiers of computing infrastructure: in-vehicle edge computing units that provide ultra-low latency processing, fog nodes typically co-located with RSUs to support intermediate computation and storage, and centralized cloud servers that offer high-capacity, resource-rich processing environments. Core functions of this layer include real-time feature extraction, object detection and classification, sensor fusion, traffic state estimation, anomaly detection, and contextual reasoning. These operations are supported by data analytics modules and distributed computing platforms that manage parallel data pipelines. Prediction and optimization engines use both model-based (e.g., Kalman filtering, traffic flow theory) and data-driven (e.g., deep learning, reinforcement learning) techniques to generate short- and long-term forecasts, such as vehicle trajectory prediction, congestion level estimation, and adaptive infrastructure scheduling. Data storage systems in this layer are responsible for real-time buffering, historical archiving, and query support. Security and fault tolerance modules detect and isolate corrupted or adversarial inputs, execute redundancy protocols, and maintain continuity of computational services under failure conditions. Additionally, digital twin models could be instantiated in this layer to simulate and monitor the real-time status of physical assets and infrastructure, thereby enabling predictive diagnostics and what-if analysis.\\

The Control Layer is responsible for translating computational outputs into executable physical actions, such as actuation of traffic signals, lane control devices, variable message signs, and vehicular subsystems. This layer integrates decision-making algorithms, emergency control subsystems, performance evaluation units, feedback mechanisms, and control logic modules. Decision-making algorithms employ a range of control strategies—including rule-based logic, model predictive control (MPC), and reinforcement learning—to generate control commands based on current and predicted system states. Emergency control subsystems monitor for critical events (e.g., crashes, infrastructure failures, inclement weather) and override normal operations to initiate corrective actions such as emergency rerouting or signal preemption. Performance evaluation units monitor key performance indicators (e.g., delay, throughput, fuel consumption, emissions) and use these metrics to tune control parameters dynamically. Feedback mechanisms support closed-loop control by continuously comparing observed outcomes against desired system objectives, and adjusting control actions accordingly. Control logic may be deployed in centralized, decentralized, or hybrid configurations to achieve a balance between coordination efficiency and actuation latency. This layer supports both time-triggered (periodic updates) and event-triggered (condition-based interventions) control paradigms, ensuring robustness and responsiveness under both normal and degraded operating conditions.\\

The Service Layer interfaces the TCPS with external users and systems, delivering high-level services built upon the outputs of the lower layers. This layer synthesizes information and provides functionality for a range of domains, including traffic management, public transportation operations, emergency response coordination, fleet and asset monitoring, and electronic toll collection. For traffic management, services include adaptive signal control, congestion monitoring, incident detection, and traveler information dissemination. Public transport services may encompass vehicle tracking, real-time arrival prediction, dynamic re-routing, and passenger load estimation. Emergency and safety services include automated crash notification, first responder coordination, hazardous condition alerts, and integration with emergency vehicle preemption systems. Fleet and asset management services support predictive maintenance, route optimization, and fuel efficiency monitoring, while tolling services may include dynamic pricing and automatic vehicle classification. These services are delivered through application programming interfaces (APIs), user dashboards, and mobile/web applications and are informed by real-time and historical data maintained in system repositories. The service layer also supports service-level analytics, compliance monitoring, and longitudinal performance evaluation to inform both operational adjustments and long-term policy-making. Importantly, the services listed are not exhaustive. The modular nature of this layer allows for the incorporation of emerging applications, such as connected pedestrian assistance, automated enforcement systems, weather-adaptive driving alerts, and environmental impact monitoring, depending on the specific deployment objectives and regulatory context.\\

Collectively, the TCPS architecture supports modularity, scalability, and interoperability. Each layer performs its designated function while exchanging information with adjacent layers through structured interfaces and data pipelines. This layered configuration enables adaptive, real-time coordination between sensing, analysis, control, and service delivery mechanisms, providing a robust foundation for the implementation of next-generation intelligent transportation systems, e.g., connected and autonomous mobility ecosystems.

\subsection{Communication in TCPS}
\label{subsec:2}

TCPS rely on a variety of communication types and protocols to connect vehicles, infrastructure, pedestrians, and backend systems. In this section, we will discuss the communication categories in TCPS. We will also discuss the TCPS communication protocols and how these protocols support use cases like cooperative driving, payment and tolling, safety messaging, and infotainment. All the communications in TCPS can be categorized into main five types:\\

\subsubsection{Vehicle-to-Vehicle (V2V)}
V2V communication refers to direct communication between vehicles, typically to exchange status information, such as position, speed, heading, and warnings. V2V forms a vehicular ad hoc network, enabling cooperative awareness among nearby cars for collision avoidance, cooperative driving (e.g., platooning), and other safety functions. The primary goal of V2V is to prevent accidents by sharing real-time data; for example, broadcasting a Basic Safety Message (BSM) ten times per second is a common approach in V2V safety systems~\cite{Twardokus2024,Arena2019}.

\subsubsection{Vehicle-to-Infrastructure (V2I)}
Communication between vehicles and roadside infrastructure, including traffic signals, RSUs, toll booths, and parking meters is referred to as V2I communication. V2I allows vehicles to receive information like signal phase and timing (SPaT) from traffic lights or to send probe data to the infrastructure. It is generally implemented using the same radio technologies as V2V (e.g., DSRC or cellular V2X), but one end is a stationary infrastructure node. Examples of V2I messages include SPaT and MAP (map data for intersections), enabling vehicles to adjust to traffic light changes, and hazard warnings broadcasting~\cite{Arena2019}.

\subsubsection{Vehicle-to-Network (V2N)}
Communication between vehicles and remote services or the internet, usually via cellular networks is known as V2N communication. In V2N, the vehicle connects to cloud servers or other internet hosts for services like traffic information, software updates, infotainment streaming, or telematics data upload. This typically involves the vehicle acting as a mobile client on 4G/5G networks to reach cloud platforms. Vehicle-to-Cloud (V2C) is a subset referring to connectivity with cloud services (for example, for over-the-air updates or remote diagnostics using standards like Diagnostics over IP). V2N communication complements direct V2V/V2I by enabling broader area coverage and backend data exchange (e.g., a car uploading sensor data to a traffic management center via cellular communication network)~\cite{Arena2019}.

\subsubsection{Vehicle-to-Pedestrian (V2P)}
V2I communication is the communication between vehicles and vulnerable road users (e.g., pedestrians and cyclists) to improve their safety. V2P may involve smartphones or wearable devices carried by pedestrians communicating with vehicles. For instance, a pedestrian’s phone could broadcast its location via an app to alert nearby connected cars. Implementations include using cellular V2X (where cell phones support the direct PC5 interface) or other wireless technologies, such as Bluetooth Low Energy, for localized alerts. V2P aims to reduce collisions by making pedestrians part of the connected ecosystem (e.g., warnings to both driver and pedestrian when a collision risk is detected)~\cite{Arena2019}.

\subsubsection{Intra-Vehicle Communications}
Intra-vehicle communications refers to the communication within the vehicle among its numerous Electronic Control Units (ECUs), sensors, and actuators. Modern vehicles contain multiple in-vehicle networks: for example, Controller Area Network (CAN) buses for engine and chassis control, Local Interconnect Network (LIN) for low-speed devices, FlexRay for high-speed and time-critical systems (e.g., x-by-wire chassis control), MOST (Media Oriented Systems Transport) for multimedia, and Automotive Ethernet for high-bandwidth data (cameras, infotainment). These wired protocols ensure different parts of a vehicle can exchange data reliably and in real-time. For instance, CAN is a robust bus used for powertrain and body control modules, while FlexRay provides deterministic timing for safety-critical controls. Intra-vehicle communications are the foundation of an in-vehicle cyber-physical system, and they interface with the V2X domain via gateway units that connect the in-vehicle network to external networks ~\cite{BenChehidaDouss2023}.

\subsection{Communication Protocols in TCPS}
\label{subsec:2.2}
TCPS relies on a variety of communication protocols across all network layers to connect vehicles, infrastructure, pedestrians, and backend systems. These range from specialized short-range vehicular networks for safety-critical exchange, to wide-area cellular links for cloud connectivity. In this section, we present an overview of major communication protocols in TCPS organized by the Open Systems Interconnection (OSI) layers~\cite{Zimmermann1980}. 

\begin{table}[ht]
\centering
\caption{Commonly Used Communication Protocols in TCPS}
% \begin{tabular}{p{3.4cm}p{8cm}}
\begin{tabular}{%
  >{\raggedright\arraybackslash}p{3.0cm}%
  >{\raggedright\arraybackslash}p{8.4cm}%
  }
\hline\\
\textbf{OSI Layer} & \textbf{Commonly Used Communication Protocols}\\\\
\hline\\
Physical \text{and} Data Link &
\textbullet\ 3GPP LTE V2X rel-14 sidelink \& 5G NR V2X sidelink – direct cellular links\\
&\textbullet\ 3GPP LTE/5G (Uu) – wide-area V2N/V2C\\\\
\hline\\
Network &
\textbullet\ WSMP – one-hop broadcast (U.S.)\\
&\textbullet\ ETSI GN – geographic multi-hop routing (EU)\\
&\textbullet\ IPv6 – conventional routing for non-safety traffic\\\\
\hline\\
Transport &
\textbullet\  BTP – lightweight, connection-less (EU)\\
&\textbullet\ UDP / TCP – standard Internet transports over IPv6\\\\
\hline\\
Application &
\textbullet\ Safety: SAE J2735 (BSM, SPaT-MAP, etc.) and ETSI CAM/DENM\\
&\textbullet\ Telematics: MQTT (publish/subscribe), CoAP, HTTPS/REST; ISO 13400 DoIP for remote diagnostics\\\\
\noalign{\smallskip}\hline
\end{tabular}
\begin{flushleft}
\textbf{Note}:
\textbf{3GPP} - 3rd Generation Partnership Project;
\textbf{LTE} - Long Term Evolution;
\textbf{5G} - Fifth-Generation;
\textbf{NR} - New Radio;
\textbf{Uu} - 3GPP device-to-base-station interface uplink/ downlink “U-u”);
\textbf{WSMP} - WAVE Short Message Protocol;
\textbf{ETSI} - European Telecommunications Standards Institute;
\textbf{GN} - GeoNetworking;
\textbf{EU} - European Union;
\textbf{BTP} - Basic Transport Protocol;
\textbf{UDP} - User Datagram Protocol;
\textbf{TCP} - Transmission Control Protocol;
\textbf{CAM} - Cooperative Awareness Message;
\textbf{DENM} - Decentralized Environmental Notification Message;
\textbf{MQTT} - Message Queuing Telemetry Transport;
\textbf{CoAP} - Constrained Application Protocol;
\textbf{HTTPS} - Hypertext Transfer Protocol Secure;
\textbf{REST} - Representational State Transfer;
\textbf{ISO} - International Organization for Standardization; and
\textbf{DoIP} - Diagnostics over Internet Protocol.
\end{flushleft}
\label{tab:tcps_layers}
\end{table}

\subsubsection{Physical And Data-link Layer Protocols}

Early TCPS field trials in North America relied on IEEE 802.11p (DSRC) operating at 5.9 GHz \cite{Bogard2017}.  More recent deployments increasingly use 3GPP C-V2X. For direct communication, LTE-V2X sidelink (PC5) has been used recently with the same 5.9 GHz band \cite{5GAA_USTrials2023}. Compared with DSRC, the LTE‐V2X sidelink (PC5) offers approximately twice the communication range, supports larger platoons under congestion, and maintains higher packet-delivery rates in urban fading conditions \cite{MolinaMasegosa2017,MolinaMasegosa2020}.  C-V2X also gains regulatory support: the 2020 FCC Report and Order reallocates most of the U.S. 5.9 GHz band to C-V2X and Wi-Fi, effectively de-prioritizing DSRC \cite{FCCRO2020}. The evolution within 3GPP demonstrates the improvement in the radio. 3GPP Release 14 introduced the basic LTE-V2X sidelink with autonomous resource selection \cite{3GPP36885}. 3GPP Release 16 added 5G NR-V2X, bringing flexible sub-carrier spacing, hybrid Automatic Repeat Request (HARQ) feedback for error correction, and Multiple-Input Multiple-Output (MIMO) antennas, while Release 17 (and ongoing Release 18 work) further reduces latency below 5 ms, increases throughput beyond 100 Mbps, and adds improved reliability modes for platoons and sensor sharing \cite{MolinaMasegosa2020,5GAA_USTrials2023}. These step-wise enhancements make C-V2X a stronger candidate both for core safety-critical applications and for the high-bandwidth sharing of sensor data (collective perception) among vehicles and infrastructure.

% These step-wise enhancements make C-V2X a stronger candidate for future connected-vehicle safety and high-bandwidth perception exchange.

\subsubsection{Network Layer Protocols}
In the U.S., the WAVE Short Message Protocol (WSMP) is used for one-hop broadcast of safety messages, allowing applications to bypass the overhead of IP and UDP headers for fast, small packets \cite{IEEE1609dot3}. In Europe, an IPv6-based GeoNetworking (GN) protocol (ETSI EN 302 636) is employed, which adds geographic addressing and multi-hop forwarding (e.g., forwarding hazard warnings along a highway) \cite{ETSI302636}. Experiments have shown that GN can propagate safety messages across multiple hops with a manageable channel load \cite{Kuhlmorgen2015}. Standard IPv6 networking is also supported in TCPS for non-safety services, enabling vehicles to use conventional routing when connected to broader networks \cite{IEEE1609dot3}.\\

\subsubsection{Transport Layer Protocols}
Above the network layer, Europe’s Basic Transport Protocol (BTP) provides a connectionless transport similar to UDP, optimized for periodic broadcast messages like Cooperative Awareness Messages (CAM) and Decentralized Environmental Notification Messages (DENM) \cite{ETSI63651}. BTP adds minimal header overhead (\textless{} 2\%) and co-exists with traditional UDP/TCP transport protocols, which are used in TCPS whenever IP connectivity is available, for instance, in cloud connectivity or infotainment traffic \cite{Kuhlmorgen2015}. This layered design allows safety-critical traffic to use lightweight transports while still supporting standard Internet protocols for other data streams \cite{ETSI63651}.\\

\subsubsection{Application Layer Protocols}
For safety-critical applications, vehicles use standardized message sets. In North America, the SAE J2735 standard defines messages, such as the BSM, SPaT, and Map Data (MAP), whereas Europe uses ETSI-defined CAM and DENM for similar purposes \cite{SAEJ2735,ETSICAMDENM}. Studies have shown that using either LTE-V2X or 802.11p can meet the typical 10 Hz broadcast rate and 100 ms latency requirement for these safety messages under moderate traffic conditions \cite{MolinaMasegosa2020}. For telematics and connected services, automotive OEMs are increasingly adopting protocols like MQTT (a lightweight publish/subscribe messaging protocol) over traditional HTTP-based REST because of MQTT’s lower communication overhead \cite{MQTTSurvey2022}. Empirical evaluations report that MQTT can reduce communication latency and bandwidth usage by roughly 10–30 \% compared with HTTP in vehicular scenarios, especially when combined with modern transport protocols, such as QUIC \cite{MQTTvsHTTP2021}. Meanwhile, ISO 13400 DoIP (Diagnostics over IP) is used for remote diagnostics to secure the connection \cite{ISO13400}. This allows maintenance or software updates on Electronic Control Units (ECUs) via the vehicle’s Ethernet gateway and cellular back-haul \cite{ISO13400}.\\

\subsection{Vulnerabilities in TCPS Communication}
\label{subsec:3}

In this section, we categorize TCPS cyber threats into broad classes that capture what the attacker aims to achieve and where the weakness lies in the communication stack. Table~\ref{tab:TCPS-Taxonomy} and the discussion that follows synthesize these works into six overlapping categories relevant to TCPS.

% \textcolor{blue}{\textbf{Figure 2 illustrates some of TCP's vulnerabilities; however, TCP has many more known vulnerabilities than those depicted in the figure.}}
% \begin{tabular}{p{1.8cm}p{4.3cm}p{5.3cm}}
\begin{table}[!t]
\centering
\caption{Classification of TCPS Communication Threats, Objectives, and Strategies}
\label{tab:TCPS-Taxonomy}
% \begin{tabular}{p{1.8cm}p{4.3cm}p{5.3cm}}
\begin{tabular}{%
  >{\raggedright\arraybackslash}p{1.8cm}%
  >{\raggedright\arraybackslash}p{4.3cm}%
  >{\raggedright\arraybackslash}p{5.3cm}%
  }
\hline\\
\textbf{Security Property} &
\textbf{Adversary Objective} &
\textbf{Attack Strategies} \\\\ \hline \\

% --- Row 1 ---------------------------------------------------------------
Confidentiality &
Extract sensitive data during communication &
% \textbullet\ Passive eavesdropping on V2X messages\\
% &&\textbullet\ Traffic analysis of toll-tag sessions\\
% &&\textbullet\ Side-channel key leakage\\\\
\begin{minipage}[t]{\linewidth}
\begin{itemize}[leftmargin=*,nosep]
  \item Passive eavesdropping on V2X messages
  \item Traffic analysis of toll-tag sessions
  \item Side-channel key leakage
\end{itemize}
\end{minipage}
\\\\ \hline \\

% --- Row 2 ---------------------------------------------------------------
Integrity &
Modify, forge, or delete legitimate messages & 
% \textbullet\ Message tampering, replay, or delay injection\\
% &&\textbullet\ False-data insertion into safety beacons\\
% &&\textbullet\ Malicious firmware tampering\\\\
\begin{minipage}[t]{\linewidth}
\begin{itemize}[leftmargin=*,nosep]
  \item Message tampering, replay, or delay injection
  \item False-data insertion into safety messages
  \item Malicious firmware tampering
\end{itemize}
\end{minipage}
\\\\ \hline \\

% --- Row 3 ---------------------------------------------------------------

Authenticity &
Impersonate an authorized TCPS entity &
% \textbullet\ Rogue RSU broadcasting forged SPaT messages\\
% &&\textbullet\ Sybil attacks with cloned certificates\\
% &&\textbullet\ GPS spoofing toward vehicles\\\\
\begin{minipage}[t]{\linewidth}
\begin{itemize}[leftmargin=*,nosep]
  \item Malicious RSU broadcasting forged SPaT message
  \item Sybil attacks with cloned certificates
  \item GPS spoofing toward vehicles
\end{itemize}
\end{minipage}
\\\\ \hline \\

% --- Row 4 ---------------------------------------------------------------

Availability &
Disrupt or block timely data delivery &
% \textbullet\ RF jamming of the 5.9 GHz band\\
% &&\textbullet\ Flooding OBUs with handshake requests\\
% &&\textbullet\ Routing misdirection\\\\
\begin{minipage}[t]{\linewidth}
\begin{itemize}[leftmargin=*,nosep]
  \item 5.9 GHz band RF jamming
  \item Flooding OBUs with handshake requests
  \item Routing misdirection
\end{itemize}
\end{minipage}
\\\\ \hline \\

% --- Row 5 ---------------------------------------------------------------

Privacy &
Correlate, track, or profile users and vehicles &
% \textbullet\ Location-tracking via fixed RFID or safety messages\\
% &&\textbullet\ Correlating toll record\\
% &&\textbullet\ Linking pseudonym certificates\\\\
\begin{minipage}[t]{\linewidth}
\begin{itemize}[leftmargin=*,nosep]
  \item Linking pseudonym certificates
  \item Correlating toll record
  \item Location-tracking via fixed RFID or safety messages
\end{itemize} 
\end{minipage}
\\\\ \hline \\

% --- Row 6 ---------------------------------------------------------------

Privilege &
Gain unauthorized execution or elevated rights &
% \textbullet\ Remote malicious code-execution on OBUs/RSUs\\
% &&\textbullet\ Remove or disable safety measures\\
% &&\textbullet\ Inject malware into the system\\\\
\begin{minipage}[t]{\linewidth}
\begin{itemize}[leftmargin=*,nosep]
  \item Execution of remote malicious code on OBUs/RSUs
  \item Removal or disablement of safety measures
  \item Injection of malware into the system
\end{itemize} 
\end{minipage} 
\\\\

\noalign{\smallskip}\hline
\end{tabular}
\begin{flushleft}
\textbf{Note}:
\textbf{V2X} - Vehicle-to-everything;
\textbf{RSU} - Roadside Unit;
\textbf{SPaT} - Signal Phase and Timing;
\textbf{GPS} - Global Positioning System;
\textbf{RF} - Radio-Frequency;
\textbf{OBU} - On Board Unit; and
\textbf{RFID} - Radio Frequency Identification.
\end{flushleft}
\end{table}

\begin{figure}[ht]
    \centering
    \includegraphics[width=\linewidth]{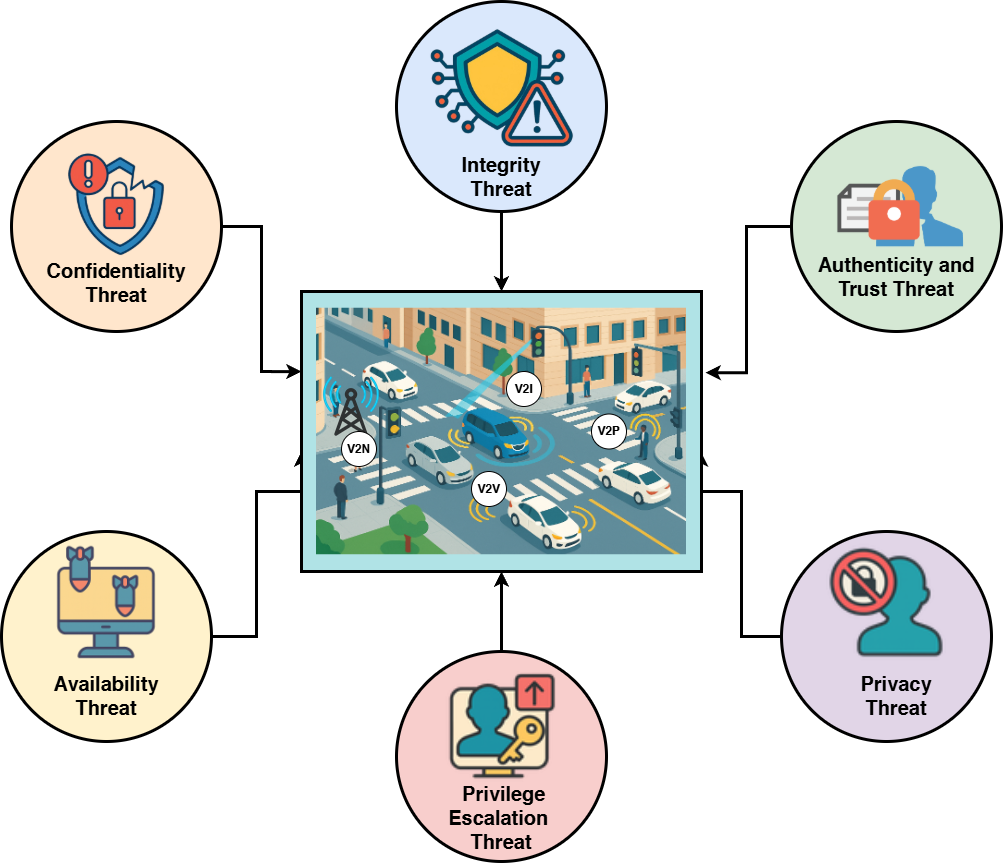}
    \caption{TCPS Vulnerabilities}
    \label{fig:TCPS-Vulnerabilities}
\end{figure}

\subsubsection{Threats Related to Confidentiality}
Adversaries who passively monitor V2X, in-vehicle, or V2N channels aim to discover sensitive information, such as location traces, driver identities, or proprietary control data.  Weak or outdated encryption keys, misconfigured TLS tunnels, or side-channel leakage from cryptographic hardware can all expose TCPS to confidentiality loss \cite{ElRewini2020}. Karim \textit{et al.} demonstrate “toll-tag sniffing,” where passive 5.9 GHz or UHF readers harvest fixed transponder IDs and reconstruct full vehicle trajectories \cite{Karim2021TollsOnly}. A study shows that using commodity laptops, it is possible to capture unencrypted BSMs and infer a driver’s route in real-time \cite{ElRewini2020}.
% Some studies have also used commodity laptops to capture unencrypted BSMs on communication links and infer a driver’s route in real-time \cite{ElRewini2020}.
Such eavesdropping needs only a roadside vantage point and highlights why strong confidentiality is essential.

\subsubsection{Threats Related to Integrity}
Safety-critical TCPS functions depend on the correctness of every packet; however, malicious alteration, selective dropping, or replay can trigger erroneous decisions—e.g., false collision warnings or incorrect lane-merge advice \cite{Kenney2011}. The “Ghost BSM” attack forges false location and speed values, causing nearby vehicles to brake or turn sharply. A study that experiments on DSRC test tracks shows a \textgreater{} 90\% success rate when message authentication is missing or replay windows are large \cite{Kenney2011}. Firmware-tamper studies further demonstrate that a single altered over-the-air update can disable automatic-braking modules, proving that integrity failures propagate quickly through in-vehicle networks \cite{Kenney2011}. Thus, robust message authentication is therefore indispensable.

\subsubsection{Threats Related to Authenticity}
Compromising the credential infrastructure or forging digital signatures allows an attacker to impersonate a vehicle, RSU, or traffic-management node \cite{Shakib2025}. Sybil attacks, where a single physical device claims many identities, can bias consensus-based applications, such as cooperative perception \cite{Brecht2018,Deka2018}. Sybil attacks have been reproduced on open testbeds by cloning a vehicle’s certificate chain and broadcasting dozens of fake identities, overwhelming cooperative-perception algorithms and degrading collision-warning accuracy by 40\% \cite{Brecht2018}. In a study, rogue RSUs replayed forged SPaT frames and tricked connected vehicles into premature acceleration at red lights \cite{Deka2018}. These examples highlight the critical need for robust authenticity in TCPS communications.

\subsubsection{Threats Related to Availability}
Denial-of-Service (DoS) attacks range from wide-band Radio Frequency (RF) jamming to protocol-level resource exhaustion that overwhelms OBUs with illegitimate handshake requests.  Availability is often the first property sacrificed in resource-constrained PQC deployments unless countermeasures, such as rate limiting and hardware acceleration, are considered \cite{Alnaseri2025}. Experiments with software-defined radios show that broadband noise can jam device-to-device communication radio signals over a 200 m stretch, raising packet loss to 95\% and disabling hazard alerts \cite{Lonc2023}. On the protocol side, flooding an on-board unit with repeated handshake requests exhausts its cryptographic engine and delays safety-message processing by more than 120 ms in stress tests \cite{Moller2016}.

\subsubsection{Threats Related to Privacy}
Even when confidentiality and authenticity are preserved, metadata can reveal personal patterns. Studies have demonstrated long-term vehicle tracking by correlating unsigned or poorly pseudonymized safety broadcasts and by harvesting fixed toll-tag IDs along road corridors \cite{Arena2019,ElRewini2020}. Combined with toll-tag scans, an attacker can build day-long location profiles even when payloads remain encrypted \cite{Karim2021TollsOnly}.

\subsubsection{Threats Related to Control/Privilege-Escalation}
Software vulnerabilities in ECUs, RSUs, or even back-end servers allow attackers to execute arbitrary code, pivot laterally, or disable safety interlocks. Unlike pure communication attacks, these often exploit implementation bugs, but weakened cryptography (e.g., broken signatures) can facilitate the initial foothold \cite{Moller2016}. High-profile incidents such as the remote Jeep Cherokee hack demonstrated full control of steering and brakes through an infotainment firmware flaw, illustrating how a single vulnerability can cascade across ECUs \cite{Moller2016}. Similar supply-chain attacks on roadside units allowed adversaries to inject malicious code that rewrote signal-phase plans, creating rolling gridlock during controlled attack simulations \cite{ElRewini2020}.\\

Mapping concrete exploits (spoofing, jamming, malware injection) onto the classes above clarifies which security goals—confidentiality, integrity, availability, privacy, or control—are at risk, and which cryptographic or systemic counter-measures are most appropriate.  The taxonomy also guides the subsequent sections: Section~\ref{sec:3}
shows how quantum computing threatens each class by breaking classical primitives, while Section~\ref{sec:4} evaluates which PQC families best restore those properties for TCPS.

\section{Classical Cryptography in TCPS Communication and Its Vulnerabilities}
\label{sec:3}

\subsection{Classical Cryptography in TCPS Communication}
Modern TCPS communications rely on a range of conventional cryptographic algorithms to fulfill security requirements. Conventional or classical security primitives fall into three functional types: public-key cryptography, symmetric-key cryptography, and cryptographic hash functions. These primitives are used for secured key establishment, data encryption, and data integrity. In the following subsections, we will discuss these cryptographic primitives in details.

\subsubsection{Symmetric-Key Cryptography }
Symmetric-key cryptography uses a single secret key shared between communicating parties for both encryption and decryption. It is valued in TCPS for its speed and efficiency in securing data streams (e.g., encrypting vehicle-to-vehicle messages or in-vehicle CAN bus traffic in real-time). Advanced Encryption Standard (AES) \cite{AES2001} is a prime example of a symmetric cipher widely used in transportation systems. AES is a block cipher, operating on fixed-size blocks with rounds of substitutions and permutations (confusion and diffusion operations). Its security rests on a large key space (e.g., $2^{128}$ possibilities for a 128-bit key) and the infeasibility of brute-force search or cryptanalysis on well-designed ciphers \cite{AES2001}. Other examples include legacy ciphers like DES/3DES \cite{DES}, which is now deprecated due to their smaller key sizes, and modern stream ciphers like ChaCha20 \cite{ChaCha20}. The underlying principle is that no efficient analytical attack is known – an adversary must essentially guess the key, a task that is computationally intractable with classical computing resources.

\subsubsection{Public-Key Cryptography}
Public-key (asymmetric) cryptography uses mathematically related key pairs: a public key, which is openly shared, and a private key, which is kept secret. It enables essential functionalities in TCPS, such as secure key exchange and digital signatures. Unlike symmetric cryptosystems, public key cryptosystems derive their security from the computational hardness of certain mathematical problems. RSA \cite{RSA}, for instance, is grounded in the difficulty of factorizing large integers, while Diffie–Hellman (DH) \cite{Diffie} key exchange and Digital Signature Algorithm (DSA) \cite{Menezes1997} rely on the discrete logarithm problem in modular arithmetic. Elliptic Curve Cryptography (ECC) \cite{Menezes1997}, used in ECDSA \cite{ECDSA} signatures and ECDH \cite{ECDH} key agreement, is based on the elliptic curve discrete logarithm problem \cite{Menezes1997}. These problems have no known efficient solution on classical computers, therefore algorithms like RSA, DH, and ECC can use moderate-size keys (e.g., 2048-bit RSA or 256-bit ECC) to achieve strong security against classical adversaries. In TCPS deployments, ECC has been especially popular due to its smaller key lengths for a given security level (e.g., 256-bit ECC key vs 3072-bit RSA for ~128-bit security), which is beneficial for resource-constrained vehicles \cite{ElRewini2020}. Thus, public-key cryptography has been instrumental for establishing trust in vehicular networks – for example, vehicles sign messages with ECDSA and exchange keys using protocols like ECIES or ECDH in order to derive shared keys and secure communications.

\subsubsection{Cryptographic Hash Functions}
Hash functions are one-way cryptographic algorithms that generate a fixed-length output, known as a hash or digest, from an input of arbitrary length. They are fundamental in TCPS for ensuring data integrity. For instance, hashes are used to verify that messages or software updates have not been tampered with, to generate pseudonymous identifiers, and as components of digital signature schemes (vehicles often sign the hash of a message rather than the full message). A good hash function, such as SHA-256 (from the SHA-2 family) or SHA-3, has strong security properties. It is designed to be collision-resistant. This means it is extremely difficult to find two different inputs that produce the same output. It is also preimage-resistant. This means it is very hard to recover the original input from the hash value \cite{SHA}. Classical hash algorithms commonly used in transportation systems include SHA-256 (widely employed in certificate generation and verification) and SHA-384/512 for higher security needs. The strength of cryptographic hashes lies in the absence of any known structure that an attacker could exploit. The best known attacks are  brute-force searches for collisions or preimages, which grow exponentially with output length (e.g., $2^{256}$ operations to brute-force a 256-bit hash preimage).\\

\subsection{Quantum Vulnerabilities of Classical Cryptography in TCPS Communication}
The rise of quantum computing poses a serious threat to traditional cryptographic algorithms \cite{Mamun2024}. Quantum algorithms can exploit mathematical computations that are infeasible for classical computers, undermining the hard problems that current security is based on \cite{Mamun2024,Twardokus2024}. In particular, Shor’s algorithm (1994) can factor large integers in polynomial time on a quantum computer. This algorithm can compute discrete logarithms in polynomial time also \cite{Shor1997}. This directly breaks the mathematical foundation of virtually all widely used public-key algorithms. A sufficiently powerful quantum adversary running Shor’s algorithm could, for example, derive an RSA private key from its public key by factoring the RSA modulus or recover an ECC private key from the public elliptic curve point by solving the elliptic curve discrete log. In practical terms, RSA, DH, DSA, and ECC become completely insecure once a quantum computer of adequate size exists—no matter how large the key, Shor’s algorithm will crack it in polynomial time \cite{Mamun2024}. This is catastrophic for TCPS security: an attacker could forge the ECDSA-based certificates and messages used in vehicular networks or decrypt sensitive data exchanges, defeating authentication and privacy protections that underpin safety-critical applications. On the other hand, for brute-force search problems, Grover’s algorithm (1996) can achieve a quadratic speed-up \cite{Grover1996}. While Grover’s algorithm does not break symmetric ciphers or hash functions outright, it significantly weakens them by halving the effective security level. In essence, a brute-force key search that takes $N$ operations classically can be done in about $\sqrt{N}$ operations on a quantum computer using Grover’s technique. For a concrete example, attacking AES-128 (with $2^{128}$ possible keys) would require on the order of $2^{128}$ trials classically, but roughly $2^{64}$ trials with Grover’s algorithm – effectively reducing AES-128 to only 64-bit security. This means that 128-bit keys, once considered amply secure, may be within reach of the future. Likewise, Grover’s algorithm can find hash preimages in $2^{n/2}$ steps for an $n$-bit hash, undermining hash functions by making it much easier to find two messages with the same digest or to invert the hash. For instance, instead of $2^{256}$ operations to brute-force a 256-bit hash like SHA-256, a quantum attacker might need only on the order of $2^{128}$ operations. \\

\begin{table}[!t]
\caption{Classical Cryptographic Types in TCPS and Security Impact of Quantum Computing}
\label{tab:Classical_Crypto_Summary}
\small
\begin{tabular}{%
  >{\raggedright\arraybackslash}p{2.1cm}%
  >{\raggedright\arraybackslash}p{2.5cm}%
  >{\raggedright\arraybackslash}p{3.8cm}%
  >{\raggedright\arraybackslash}p{3.0cm}}
\noalign{\smallskip}\svhline\noalign{\smallskip}\\
\textbf{Cryptographic Type} & \textbf{Purpose in TCPS} & \textbf{Quantum Vulnerability} & \textbf{Security Against Quantum Computer} \\\\
\noalign{\smallskip}\svhline\noalign{\smallskip}\\

Symmetric Key & Encryption & Grover’s algorithm gives a quadratic speed-up for key search, effectively reducing the security margin & Secure with larger key sizes (at least double in size)  \\\\ \hline \\

Asymmetric Key &  Encryption, Digital Signature, and Key Exchange & Shor's algorithm can extract the private key from public primitives in polynomial time& Not Secure when a large enough quantum computer emerges\\\\ \hline \\

Hash Functions &  Hashing & Grover’s algorithm reduces pre-image/collision resistance, halving effective security bits & Secure with larger output size \\\\
\noalign{\smallskip}\svhline\noalign{\smallskip}
\end{tabular}
\end{table}

Table~\ref{tab:Classical_Crypto_Summary} distils the quantitative discussion above. Quantum computing threatens to undermine all three pillars of classical cryptography. Public-key schemes (the basis for most authentication and key exchange in TCPS) would be rendered obsolete by Shor’s algorithm, and symmetric ciphers and hashes would have their security levels drastically reduced by Grover’s algorithm. An attacker with a quantum computer could impersonate vehicles or infrastructure by forging digital signatures, decrypt confidential V2X communications, and find preimages in safety message hashes, wreaking havoc on transportation security. While increasing symmetric key sizes or hash lengths can somewhat patch the symmetric-key and hashing schemes (e.g., using AES-256 instead of 128, or SHA-512 instead of 256, to restore classical-level security under Grover’s model), there is no practical remedy for the loss of security in RSA/ECC aside from transitioning to new quantum-resistant algorithms. This stark reality is driving the development of post-quantum cryptography (PQC) – novel cryptographic schemes based on alternative hard problems believed to withstand quantum attacks. Adopting PQC in TCPS is essential to ensure long-term security and resilience of vehicular communications in the quantum era \cite{ElRewini2020}.
\section{Post-Quantum Cryptography (PQC) in Securing TCPS Communication}
\label{sec:4}

\subsection{PQC Overview}
\label{subsec:5}
Post-quantum cryptography is a field focused on developing cryptographic algorithms based on mathematical problems that are expected to remain secure against quantum attacks. These problems are also considered computationally difficult for classical computers \cite{NIST8105}. Unlike quantum cryptography, which relies on quantum physics for tasks such as quantum key distribution (QKD), PQC algorithms operate on standard digital hardware while remaining secure in the presence of quantum-capable adversaries. The primary goal of PQC is cryptographic agility for the quantum era—developing drop-in replacements for RSA/ECC and other vulnerable schemes so that our communications remain secure in the face of quantum breakthroughs \cite{NIST8105}. NIST has spearheaded a global effort to standardize and evaluate PQC algorithms. Starting in 2016, NIST organized an open competition-like process, attracting submissions from researchers worldwide \cite{NISTPQCProject}. Over multiple rounds of rigorous cryptanalysis and performance evaluation, the candidates were winnowed down: 69 algorithms in Round 1 were reduced to 26 in Round 2, then 15 finalists in Round 3 \cite{NIST8240,NIST8309,NIST8413}. In July 2022, NIST announced its initial selections for post-quantum cryptographic standards. One key encapsulation mechanism based on lattice cryptography, known as CRYSTALS-Kyber \cite{Kyber}, was selected as a draft standard. A lattice-based digital signature algorithm, CRYSTALS-Dilithium \cite{Dilithium}, was also chosen. In addition, two other digital signature schemes—FALCON \cite{Falcon} and SPHINCS+ \cite{SPHINCS+}—were named as finalists for further evaluation \cite{NISTPQCProject}. After an additional analysis period, in August 2024, NIST officially standardized three PQC schemes: Kyber for public-key encryption/KEM, and Dilithium and SPHINCS+ for digital signatures \cite{NISTPQCProject}. These have been assigned draft FIPS standards (FIPS 203, 204, 205, respectively) and are the first generation of quantum-resistant crypto for general use \cite{NIST2024FIPS}. Recently, in March 11, 2025, NIST standardized HQC as the fifth post-quantum encryption algorithm \cite{HQC}.

\subsection{PQC Algorithm Families}
\label{subsec:6}

The diversity of mathematical approaches in PQC is significantly greater than in classical cryptography, which was dominated by a few intractable problems, such as factoring prime numbers and discrete logarithmic problems. As presented below, five primary families of post-quantum algorithms have emerged \cite{NIST8105}: 

\subsubsection{Lattice-Based Cryptography}
Lattice-based cryptography uses problems based on high-dimensional lattices, such as the Shortest Vector Problem (SVP) \cite{SVP} or Learning With Errors (LWE) \cite{LWE} problem. These problems are based on identifying compact solutions within complex lattice structures or solving linear equations with added noise. Both tasks are considered infeasible for classical and quantum computers \cite{Mamun2024}. Lattice-based schemes have arguably become the front-runners of PQC due to their strong security and efficiency. For example, CRYSTALS-Kyber \cite{Kyber} for KEM, and CRYSTALS-Dilithium \cite{Dilithium}  for digital signature, all these NIST-standardized algorithms are lattice-based, relying on structured LWE problems over module lattices \cite{NISTPQCProject}. Advantages of lattice cryptography include relatively small per-operation computation and versatility, as the same mathematical functions can be used for encryption-decryption, key exchange, and digital signatures \cite{KyberSpec}. Key sizes and ciphertexts are moderate, and operations are fast with simple arithmetic \cite{KyberSpec}. However, lattice schemes have larger signatures or ciphertexts than RSA and ECC \cite{Lonc2023}, require randomness sampling \cite{KyberSpec}, and some schemes can be more susceptible to side-channel attacks \cite{Primas2017}. Nonetheless, lattices are currently the leading PQC category for general use, with decades of research supporting their hardness.

\subsubsection{Code-Based Cryptography}
Code-based cryptography relies on the difficulty of correcting random errors in encoded messages. In this setting, the original encoding structure is unknown. Solving this problem is widely believed to be NP-hard \cite{May2011}. The classic example is the McEliece cryptosystem (1978), which uses random binary Goppa codes \cite{McEliece}. When encrypting, the sender encodes the plaintext as a codeword of the public code and then deliberately flips a small, random set of bits. Anyone lacking the secret trapdoor, which is the private key that reveals the structure of the underlying code, must solve the generic decoding problem to recover the message—an operation regarded as computationally infeasible \cite{McEliece}. Code-based schemes have very large public keys. Classic McEliece’s public key can be several hundred kilobytes \cite{Lonc2023}, but extremely fast encryption and decryption, and a long track records of resisting attacks \cite{Mamun2024}. Classic McEliece is a 4th-round NIST finalist valued for its conservative security \cite{NIST8545}. Other structured code-based proposals, such as HQC and BIKE, were explored to reduce key sizes, although BIKE was later excluded from the NIST process due to cryptanalytic concerns \cite{nosouhi2023weak, NIST8545}. The strengths of code-based crypto are its simplicity and confidence. The obvious weakness is impractical key sizes for bandwidth- or memory-constrained applications – distributing a 0.5 MB public key to every vehicle might strain V2X channels \cite{Lonc2023}. Nonetheless, for applications like firmware signing or infrastructure where keys can be stored, code-based signatures and encryption are viable.

\subsubsection{Hash-Based Cryptography}
Hash-based cryptography depends solely on cryptographic hash functions for its security. Digital signatures (and related primitives) are built directly from these hashes, so their security reduces to the underlying hash function's resistance to collisions and pre-image attacks \cite{Mamun2024,Brassard1997}. Hash-based signatures like XMSS \cite{XMSS} and SPHINCS+ \cite{SPHINCS+} generate signatures by revealing parts of a large pool of one-time hash outputs. They are extremely robust – even a quantum computer cannot do much better than brute-force guessing to break a hash, and rely on the well-understood security of hashes \cite{Brassard1997}. SPHINCS+, now standardized by NIST, is a stateless hash-based signature scheme. It provides strong security while relying on minimal cryptographic assumptions \cite{SPHINCS+}. Its main trade-offs include very large signature sizes, often in the tens of kilobytes, and slower signing and verification times when compared to lattice-based or code-based cryptographic schemes \cite{Lonc2023}. Hash-based signatures also often impose restrictions on how many times a single key can be securely used for signing. For example, stateful schemes like XMSS depend on proper handling of internal state, since reusing one-time keys can compromise security. In TCPS, hash-based crypto might be suitable for signing software updates or other infrequent operations where size is less critical than security longevity. The lack of reliance on any algebraic structure is a plus, as no known structural attacks have been discovered so far, but the performance and size must be managed. For instance, sending a 40 KB SPHINCS+ signature over a vehicular network that transmits 10 messages per second is not feasible; thus, hash-based methods might be reserved for higher-level assurances like certificate authority signatures that are infrequently verified by vehicles.

\subsubsection{Multivariate Polynomial Cryptography}
The difficulty of multivariate polynomial cryptography is based on the Multivariate Quadratic (MQ) problem. The MQ problem refers to solving systems of multivariate quadratic equations over finite fields,  which is NP-complete \cite{Bellini2022}. These schemes produce typically digital signatures like Rainbow \cite{Rainbow} and GeMSS \cite{GeMSS} by constructing trapdoor functions from multivariate equations \cite{Bellini2022}. They offered the advantage of very short signatures and relatively fast signing speeds, but this came at the expense of large public keys and slower verification times. However, this category has suffered major setbacks: both Rainbow and GeMSS, which were once NIST finalists, were broken by cryptanalysis in 2020–2022 \cite{Tao2021,Tateiwa2024}. Algebraic attacks managed to either recover the secret key or significantly undermine the claimed security of these schemes. As a result, none of the multivariate cryptographic schemes progressed to the final round of the NIST standardization process. The multivariate approach is still of academic interest (and some niche systems might still consider it), but at present, it appears less promising due to the difficulty of constructing secure trapdoors that resist the onslaught of algebraic attack techniques. The strength of multivariate schemes was efficiency and small signatures/keys; the weakness proved to be the unintended algebraic structure that attackers could exploit \cite{Mamun2024}. Further research is needed to determine if any multivariate construction can be made robust enough for standardization.

\subsubsection{Isogeny-Based Cryptography}
Isogeny-based cryptography relies on problems from algebraic geometry, specifically finding isogenies between elliptic curves. Its security relies on the Supersingular Isogeny Diffie–Hellman (SIDH) problem: having two supersingular elliptic curves that are connected by an isogeny, an attacker has to recover that hidden isogeny \cite{Galbraith2018}. The main allure of isogeny-based schemes was the extremely small key sizes. For instance, Supersingular Isogeny Key Encapsulation (SIKE) has public keys on the order of only 264 bytes, much smaller than lattice or code keys \cite{SIKE}. This would be ideal for bandwidth-limited V2X applications. However, isogeny crypto received a huge blow in 2022 when a breakthrough classical attack using advanced math but not requiring a quantum computer was devised by Castryck \textit{et al.}, effectively breaking SIDH for the parameter sizes used in SIKE \cite{Castryck2022}. The SIKE team promptly announced the attack, and SIKE was dropped from NIST consideration. At present, isogeny-based cryptography is not standardized due to this vulnerability \cite{Mamun2024}. It remains an important research area – and variants might be strengthened or other isogeny problems might prove secure – but until a quantum-safe and classical-safe isogeny scheme is found, this family is on hold. The advantage of isogeny schemes is ultra-compact keys; however, the disadvantage is that the known schemes turned out to be insecure against classical attacks (let alone quantum). \\
% Future research (possibly integrating isogeny ideas into hybrid schemes) could revive interest if new hard problems are identified.\\

\begin{table}[!t]
\caption{Post-quantum algorithm families and TCPS use case(s)}
\label{tab:PQCFamiliesTCPS}
\begin{tabular}{%
  >{\raggedright\arraybackslash}p{1.0cm}%
  >{\raggedright\arraybackslash}p{1.3cm}%
  >{\raggedright\arraybackslash}p{1.4cm}%
  >{\raggedright\arraybackslash}p{2.4cm}%
  >{\raggedright\arraybackslash}p{2.4cm}%
  >{\raggedright\arraybackslash}p{2.9cm}}
\hline\noalign{\smallskip}\\
\textbf{PQC Family} & \textbf{Core Mathematical Problem} & \textbf{Examples} &
\textbf{Strengths} & \textbf{Limitations} & \textbf{TCPS Use Case(s)} \\\\
\hline\\
Lattice &
SVP \& LWE &
\begin{minipage}[t]{\linewidth}
Kyber, Dilithium, Falcon
\end{minipage}
&
\begin{minipage}[t]{\linewidth}
\begin{itemize}[leftmargin=*,nosep]
  \item Fast
  \item Same implementation for KEM \& signature
  \item Hardware acceleration is possible
\end{itemize}
\end{minipage}
&
\begin{minipage}[t]{\linewidth}
Large ciphertext and signature size
\end{minipage}
&
\begin{minipage}[t]{\linewidth}
Practical for V2V/V2I safety messages and certificate provisioning
\end{minipage}
\\\\
\hline\\
Code &
Syndrome decoding &
\begin{minipage}[t]{\linewidth}
Classic McEliece, HQC, BIKE
\end{minipage}
&
\begin{minipage}[t]{\linewidth}
\begin{itemize}[leftmargin=*,nosep]
  \item Long security record
  \item Fast encryption and decryption
\end{itemize}
\end{minipage}
&
\begin{minipage}[t]{\linewidth}
Very large public key
\end{minipage}
&
\begin{minipage}[t]{\linewidth}
Best for backend authorities, servers, and during heavy firmware updates
\end{minipage}
\\\\
\hline\\
Hash &
Hash pre-image &
\begin{minipage}[t]{\linewidth}
SPHINCS$+$, XMSS
\end{minipage}
&
\begin{minipage}[t]{\linewidth}
Few assumptions and robust security
\end{minipage}
&
\begin{minipage}[t]{\linewidth}
\begin{itemize}[leftmargin=*,nosep]
  \item Large signature size
  \item Slower signature verification
\end{itemize}
\end{minipage}
&
\begin{minipage}[t]{\linewidth}
Root certificates, Certificate Revocation List (CRL), and firmware signing, where size is less critical
\end{minipage}
\\\\
\hline\\
Multi-variate &
MQ equations &
\begin{minipage}[t]{\linewidth}
Rainbow, GeMSS
\end{minipage}
&
\begin{minipage}[t]{\linewidth}
\begin{itemize}[leftmargin=*,nosep]
  \item Very short signature
  \item Fast signing
\end{itemize}
\end{minipage}
&
\begin{minipage}[t]{\linewidth}
Recent cryptanalysis has broken this scheme
\end{minipage}
&
\begin{minipage}[t]{\linewidth}
Research only—could shrink V2V frames if a secure scheme is found
\end{minipage}
\\\\
\hline\\
Isogeny &
Super-singular isogeny &
\begin{minipage}[t]{\linewidth}
SIKE
\end{minipage}
&
\begin{minipage}[t]{\linewidth}
Ultra small keys
\end{minipage}
&
\begin{minipage}[t]{\linewidth}
Recent cryptanalysis has broken this scheme
\end{minipage}
&
\begin{minipage}[t]{\linewidth}
Future option for ultra-low-bandwidth sensors; not ready to deploy now
\end{minipage}
\\
\noalign{\smallskip}\hline
\end{tabular}
\begin{flushleft}
\textbf{Note:}
\textbf{SVP} - Shortest Vector Problem; 
\textbf{LWE} - Learning With Errors; 
\textbf{KEM} - Key Encapsulation Mechanism; 
\textbf{V2V} - Vehicle-to-vehicle; 
\textbf{V2I} - Vehicle-to-infrastructure; and 
\textbf{MQ} - Multivariate Quadratic. 
\end{flushleft}
\end{table}

In summary, lattice-based and code-based cryptography currently lead the field for practical PQC, with hash-based signatures providing a complementary option for specialized needs. We emphasize the trade-offs: lattice schemes (Kyber, Dilithium, Falcon) have modest key and signature sizes (a few kB at most) and are reasonably fast – making them suitable for most TCPS applications. Code-based schemes like Classic McEliece have huge keys (hundreds of kB) but very small ciphertexts and high speed, potentially viable for one-time setups or systems that can handle large static keys. For example, device manufacturers could pre-load keys in the device.  Hash-based schemes, such as SPHINCS\(+\), have large outputs and slower performance, so they may be reserved for high-assurance back-end processes rather than routine V2V messaging.  Multivariate and isogeny schemes currently have no standard, but their past candidates illustrate either performance wins negated by weak security (multivariate) or size wins negated by a critical break (isogeny). Table~\ref{tab:PQCFamiliesTCPS} presents these five families side by side, summarizing the hard problems they rely on, representative algorithms, key strengths and limitations, how each family fits typical TCPS links, and their current NIST status. This overview enables designers to see at a glance which post-quantum options best match bandwidth, latency, and deployment constraints in adequate TCPS.

\subsection{PQC Integration Challenges in TCPS}
\label{subsec:7}

An important aspect to consider is the computing resources and timing requirements of PQC algorithms when deployed in transportation systems. TCPS imposes strict real-time constraints – for example, safety messages in V2V communication often must be verified within 100 ms or less \cite{Twardokus2024}. Any cryptographic operation that significantly increases latency or message size can disrupt the system’s performance \cite{Lonc2023}. Early evaluations have shown both promise and challenges. In an experimental study, the NIST-standard Kyber KEM was integrated into a TCPS setting to secure V2X message exchanges. The results showed that Kyber performed well on high-bandwidth, low-latency wired links (e.g., securing Ethernet connections between control centers) without noticeable issues \cite{Mamun2024}. However, on latency-sensitive wireless links (such as V2V safety message broadcasts), the added overhead of PQC—larger keys and more computation—introduced difficulties in consistently meeting the tight timing requirements. For instance, the encapsulated ciphertext in a Kyber-based session might be larger than an equivalent ECDH exchange, affecting packet loss on DSRC/5G channels, and the cryptographic processing might eat into the 100 ms budget for safety applications. Similarly, replacing ECDSA with Dilithium signatures in V2X could increase message size and verification time; if a vehicle needs to verify 100 signatures per second from nearby cars, a heavier signature scheme could strain its CPU \cite{Lonc2023}. \\

These observations highlight a critical reality: integrating PQC into TCPS requires careful tailoring to meet stringent performance, bandwidth, and latency constraints. Achieving this integration demands a multifaceted approach, including parameter tuning, adoption of lightweight PQC variants, and the use of hybrid mechanisms that combine classical and PQC algorithms (e.g., dual signatures or hybrid certificates) \cite{Mamun2024}. However, identifying the optimal cryptographic configuration under dynamic operating conditions is a non-trivial challenge. This is precisely where AI can play a transformative role—enabling adaptive, context-aware selection and deployment of PQC mechanisms in real time. The role of AI in orchestrating this transition will be explored in detail in Section~\ref{sec:5}.

% \textcolor{red}{\textbf{It would be beneficial to include: 1) a performance comparison table for classical, PQC, and hybrid implementations on OBU/RSU hardware; 2) a detailed discussion on SCMS migration, addressing certificate formats, key management life cycle, and backward compatibility strategies; 3) detailed AI-switching criteria based on threat levels and resource conditions..}}\\

% \textcolor{teal}{\textbf{1- Please consider adding a table quantifying message sizes, handshake costs, and CPU/energy for representative classical, PQC, and hybrid stacks on OBU/RSU-class hardware. 2- Please outline migration implications for credentialing (e.g., SCMS) and key lifecycles under the finalized NIST PQC standards and the new backup KEM, and specify when AI should switch between hybrid and pure-PQC modes.}}\\

\section{AI-Driven PQC in TCPS}
\label{sec:5}
% \textcolor{blue}{\textbf{As the primary focus of this chapter is Post-Quantum Cryptography (PQC), the discussion on Post-Quantum Vulnerabilities (PQV) should be introduced earlier than Section 5.}}\\
% \textcolor{blue}{\textbf{With a stronger emphasis on AI-driven PQC, this chapter will be significantly improved. Please improve organization to improve the flow.}}\\
Deploying post-quantum cryptography in TCPS at scale will require overcoming challenges in algorithm selection, system integration, and real-time operation. Artificial intelligence and machine learning techniques can serve as powerful enablers in this transition by intelligently automating decisions and enhancing security beyond what deterministic algorithms can achieve. In this section, we outline how AI-driven approaches can support PQC deployment in TCPS and improve cyber-resilience. Several key opportunities for AI in this domain include: adaptive algorithm selection, threat prediction, resource optimization, anomaly detection, and managing hybrid cryptographic models.\\

\subsection{Adaptive Algorithm Selection and Cryptographic Agility}
Some of the recent research proposes ML-driven frameworks that classify data or network context and then vary cryptographic strength on the fly, balancing security with throughput. Kumar \textit{et al}. demonstrated a study on fog-computing security that a KNN-based controller can dynamically switch between AES-only encryption and a hyrid ECC-AES encryption based on data sensitivity, thereby reducing overhead while preserving confidentiality \cite{Kumar2025Fog}. Reinforcement Q-learning has likewise been used to scale encryption levels in wireless-sensor networks, reducing energy consumption 30 \% without sacrificing packet-delivery ratio~\cite{Premakumari2025}. Similar adaptive-encryption ideas appear in ~\cite{Singh2025} that uses GANs and genetic algorithms to evolve protocol choices against side-channel leakage. Collectively, these works show how AI can optimize “what to run, when” for PQC and legacy primitives in dynamic TCPS environments.\\

TCPS are heterogeneous – they encompass a range of devices (from high-power roadside servers to low-power sensors) and network conditions (high-speed 5G links, congested DSRC channels, etc.)~\cite{He2022,ElRewini2020}. No single PQC algorithm is optimal for all scenarios~\cite{Vidakovic2023}; for example, a resource-constrained sensor might handle a small hash-based signature better than a large lattice key exchange, whereas a road-side unit with more bandwidth might do the opposite~\cite{Suhail2021}. AI can be used to develop intelligent cryptographic agility frameworks that choose the most suitable algorithm and parameters for each context. A machine learning model could be trained on parameters like message size, required latency, node CPU load, and channel conditions to decide whether to use, say, Dilithium or Falcon for a given signature, or whether to engage a hybrid RSA+Kyber handshake or a pure Kyber handshake for key exchange. By learning from past performance data, such an AI system would effectively tune the cryptographic scheme in real-time to ensure both security and efficiency. Ali \textit{et al}. demonstrate in \cite{Ali2025} that a reinforcement-learning engine with meta-learning can monitor device load, battery state, latency budget, and threat score to choose the most appropriate post-quantum primitive in real time, showing how AI enables true cryptographic agility in heterogeneous TCPS. This kind of AI-driven selection is especially valuable as new PQC options emerge—the AI can continuously adapt, unlike static configurations. This continual learning loop preserves cryptographic agility across the entire TCPS, ensuring every node always runs the lightest algorithm that meets its current latency, bandwidth, and security targets.
\subsection{Threat Prediction and Proactive Defense}
AI excels at analyzing large datasets to find patterns and predict future events. In TCPS security, an AI model can ingest diverse threat-intelligence feeds—ranging from reports on quantum-computing advances and logs of real-world encryption-breach attempts to abnormal patterns in in-vehicle network traffic \cite{Althunayyan2025}—and use that data to predict emerging attacks or latent vulnerabilities. If new research indicates that a chosen PQC primitive is losing strength, the AI controller can flag the issue and initiate a phased migration to an alternate algorithm \cite{NIST8547}. Recent forecasting studies estimate that fault-tolerant quantum computers capable of factoring RSA-2048 are unlikely (< 5 \% probability) to appear before 2039, although some industry scenarios place a non-negligible risk window as early as 2030 \cite{SecureworksQDay,Sevilla2020}. Guided by these forecasts, TCPS operators can follow NIST crypto-agility playbooks to rotate keys or swap in quantum-resistant suites well before the threat emerges, increasing modulus sizes or activating PQC where needed \cite{NCCoE2024}. Reinforcement-learning controllers can automate that response, tightening security policies (e.g., mandatory PQC despite the added latency) when local threat scores spike and relaxing them when conditions normalise, thereby preserving both safety and performance \cite{Premakumari2025}. In other words, when anomaly scores spike in a specific road corridor or time window, the AI policy engine automatically locks that segment into a “PQC-only, extended-key” profile—accepting the added latency until the risk subsides \cite{Premakumari2025}. In essence, AI delivers a risk-aware, adaptive security posture for TCPS that balances safety and performance as conditions evolve \cite{Ali2025,Premakumari2025}.
\subsection{Resource Optimization for PQC Operations}
Large post-quantum keys and signatures can saturate low-bandwidth vehicular links and strain embedded processors; cross-platform and embedded-hardware studies of Kyber, Dilithium and McEliece document substantial memory, bandwidth and energy overheads \cite{Alnaseri2025,Pursche2024,Atkins2021,Keysight2025}. An AI-driven scheduler might mitigate this high requirements by allocating cryptographic workloads intelligently or selecting lighter parameter sets under tight deadlines. Kumar \textit{et al.} showed in a fog-computing experiment that a KNN-based controller can switch between AES-only and a hybrid ECC-AES based on data sensitivity, lowering overhead while preserving confidentiality \cite{Kumar2025Fog}. Reinforcement-learning frameworks similarly scale cryptographic strength in wireless-sensor networks, and surveys of deep-RL schedulers highlight their applicability to edge and cloud resource management \cite{Premakumari2025,Gu2025}.Performance analyses note that lattice algorithms demand far more CPU cycles and memory, while code-based keys can exceed a megabyte—constraints that challenge vehicle and roadside hardware \cite{Commey2025}. Deep-RL offloading frameworks for vehicular edge-cloud networks decide when to shift heavy PQC computations to roadside servers, balancing latency and load \cite{Almuseelem2025,Shi2023}. Machine-learning techniques, such as Bayesian optimization and genetic search, tune lattice parameters to find security–speed sweet spots tailored to each processor \cite{Ajax2025}. On the other hand, Generative-AI models that synthesise realistic V2X traffic traces enable automatic protocol tuning, improving both latency and security margins under dynamic road conditions \cite{Zhang2024}. Bhatia \textit{et al.} demonstrate that resource-aware deep-RL schedulers can both trim Kyber and Dilithium handshake times on Raspberry-Pi-class boards and keep those devices within tight energy budgets \cite{Bhatia2019}.

\subsection{Anomaly and Intrusion Detection}
AI-based anomaly detection is already a growing field in vehicle cybersecurity \cite{Venkatasamy2024,Althunayyan2025,Jayasri}.
As PQC is deployed, adversaries are expected to pivot from breaking encryption to exploiting side-channel leaks and implementation flaws \cite{Ravi2023,Ravi2021NIST}. Deep-learning models trained on vehicular traffic have demonstrated real-time detection of replay, spoofing and handshake anomalies in V2X message flows. For example, Venkatasamy et al. built a machine-learning intrusion-detection system that flags suspicious repetition patterns, while Jayasri \textit{et al.} achieved high accuracy with an LSTM-based detector for injection and DoS attacks \cite{Venkatasamy2024,Jayasri}. AI can extend this capability to side-channel monitoring: lattice-based schemes like Kyber and Frodo have been shown to leak exploitable timing or power information \cite{Ravi2023,Ravi2021NIST,Ajax2025Side}, and AI-driven countermeasures can classify leakage traces and trigger adaptive protections before an attacker recovers secret keys \cite{Ajax2025Side,Hoang2024}. Research by Hoang \textit{et al}. demonstrates that a convolutional-neural-network attack can recover a full Kyber private key with only 50 power traces, underscoring the need for continuous monitoring. In parallel, reinforcement-learning schedulers and generative-AI traffic synthesizers can selectively raise security levels—enforcing PQC even at the cost of extra latency—only during periods of elevated risk, thereby balancing safety and performance \cite{Premakumari2025,David2024}. Collectively, these findings indicate that intelligent anomaly detectors provide a second line of defence for TCPS, catching novel tactics that attempt to bypass or undermine PQC implementations \cite{Althunayyan2025,Ejaz2025}.
\subsection{Hybrid Cryptographic Models and AI Orchestration}
NIST Internal Report (IR) 8547’s initial public draft explicitly endorses “hybrid solutions” that pair an approved classical algorithm with a newly standardised PQC algorithm and insists that governance policies stipulate when the legacy component must be retired \cite{NIST8547}. Early prototypes already realise these requirements: ECDSA-Dilithium security-key firmware, TLS/SSH test-beds with composite KEMs, and Microsoft-backed proposals for ECDH-Kyber handshakes illustrate how dual artifacts are carried in real traffic \cite{Ghinea2022,Stebila2019,PQHybrid2024}. Such hybrids nearly double ciphertext or certificate size and add extra signing and verification steps, creating bandwidth and CPU challenges for roadside units and embedded vehicle controllers \cite{Stebila2019,GSMA2024}. AI techniques can automate those tradeoffs between different requirements: reinforcement-learning agents have been shown to rotate Kyber or Dilithium keys on demand, switch between hybrid and pure-PQC paths, and cut management overhead compared with static policies \cite{Ajax2025RL,Bamigboye2025}. Survey work on ML-assisted optimisation confirms that intelligent parameter tuning can shrink key sizes and reduce handshake latency without manual intervention \cite{Mmaduekwe2025}. In a TCPS deployment, an AI controller could track upgrade status across vehicles and RSUs, decide when to transmit a single Dilithium signature versus an ECDSA + Dilithium pair \cite{GSMA2024,NIST8547}.\\\\

\begin{table}[!t]
\caption{AI-driven techniques to support PQC deployment in TCPS}
\label{tab:AITechniques}
% \begin{tabular}{p{2.1cm}p{3.0cm}p{2.8cm}p{3.5cm}}
\begin{tabular}{%
  >{\raggedright\arraybackslash}p{2.1cm}%
  >{\raggedright\arraybackslash}p{3.0cm}%
  >{\raggedright\arraybackslash}p{2.8cm}%
  >{\raggedright\arraybackslash}p{3.5cm}}
\noalign{\smallskip}\svhline\noalign{\smallskip}\\
\textbf{Functionality} & \textbf{Purpose} & \textbf{AI/ ML Methods} & \textbf{Impact(s) on TCPS} \\\\
\noalign{\smallskip}\svhline\noalign{\smallskip}\\

Adaptive selection & Context-aware choice of cryptographic scheme & KNN, RL policy & Meets sub-100 ms latency, saves bandwidth \\\\

\hline \\

Threat prediction & Forecast quantum-related cyber risk & Forecast models, anomaly scores & Early migration, risk-responsive protection \\\\

\hline \\

Resource optimization & Keep PQC operations within CPU and battery power limits & RL off-loading, Bayesian tuning & less computational energy and lower communication delay \\\\

\hline \\

Anomaly detection & Catch spoofing, side channel leakage & CNN/LSTM classifiers & Blocks forgery or key-leak attempts \\\\

\hline \\

Hybrid orchestration & Manage hybrid classical-PQC implementations & RL key-lifetime manager, parameter search & Reduces certificate size and CPU usage. \\\\
\noalign{\smallskip}\svhline\noalign{\smallskip}
\end{tabular}
\begin{flushleft}
\textbf{Note}:
\textbf{KNN} - k-Nearest Neighbors;
\textbf{RL} - Reinforcement Learning,\;
\textbf{CNN} - Convolutional Neural Network; and
\textbf{LSTM} - Long Short-Term Memory.
\end{flushleft}
\end{table}

These applications of AI in PQC for TCPS suggest that by combining PQC’s quantum resistance with AI’s ability to dynamically improve based on data, AI-enabled cryptographic frameworks not only withstand quantum attacks but also self-optimize against new threats over time. Of course, these integrations are not without challenges: the computational overhead of AI and PQC together can be significant, and careful consideration must be given to ensure that adding AI does not introduce new vulnerabilities (for example, adversaries tricking the ML model). Nonetheless, the synergy of AI and PQC holds great promise for future-proofing TCPS communications. In summary, AI-driven PQC can make TCPS not only quantum-safe but also more intelligent and more autonomous in maintaining their security, aligning with the broader trend of autonomy in transportation. Table~\ref{tab:AITechniques} gives a concise overview of these AI techniques, showing for each function the purpose, common methods, literature examples, and measured benefits. Designers can use the table as a quick checklist when selecting which AI module to pair with a given PQC deployment scenario.

\section{Future Research Direction}
\label{sec:6}
Transportation Cyber-Physical Systems are at the cusp of a significant security evolution. In this chapter, we have discussed how the looming reality of quantum computers necessitates a transition from classical cryptographic algorithms to post-quantum cryptographic algorithms to ensure the long-term security of TCPS communications. We began by highlighting the crucial role of cryptography in TCPS—protecting everything from vehicle coordination messages to toll payments—and reviewed the numerous vulnerabilities that could be exploited if these protections fail. We found that quantum algorithms such as Shor's and Grover's completely break the security foundations of RSA, ECC, and other classical cryptographic algorithms. This leaves TCPS communications vulnerable to impersonation, eavesdropping, and manipulation by adversaries with quantum capabilities. This motivates an urgent deployment of PQC algorithms, which come from diverse families (e.g., lattice, code, and hash), each offering quantum resistance under different trade-offs. The NIST PQC standardization program has provided initial standards (i.e., Kyber, Dilithium, SPHINCS+) that TCPS stakeholders can begin adopting. However, integrating these into real-world transportation systems poses challenges in meeting strict latency, bandwidth, and resource constraints.\\

We then explored the idea of leveraging AI-driven approaches to facilitate and strengthen the PQC deployment in TCPS. AI can act as a force multiplier for security: machine learning algorithms can dynamically choose cryptographic parameters, predict and preempt attacks, optimize performance, detect anomalies, and coordinate complex hybrid cryptosystems in ways that static configurations cannot. By infusing intelligence into the security stack, TCPS can become cyber-resilient – capable of adapting to new threats on the fly, and optimizing security so that it does not impede the functionality of the system. The concept of AI-driven PQC thus represents a convergence of two advanced fields – AI and PQC – with the goal of creating transportation networks that are secure against both current and future threats. Looking ahead, several promising research and development directions emerge:\\

\begin{enumerate}
\item \textbf{Lightweight PQC for TCPS:} There is a need for cryptographic schemes tailored to the constrained environments of vehicles and IoT devices in transportation. This could involve designing new PQC algorithms (or variants of existing ones) that significantly reduce key sizes and computational load, specifically for scenarios like V2X, where packets are small and frequent. Techniques like key compression, hardware acceleration, and protocol-level optimizations also fall under this umbrella. The goal is to minimize the PQC overhead so that upgrading security does not degrade performance. As noted earlier, experiments highlight that standard PQC solutions might struggle with the 10 ms latency requirements of V2V safety messages \cite{Lonc2023}, so lightweight PQC research aims to bridge that gap. This could also extend to energy-efficient PQC, ensuring that battery-powered devices (like sensors or electric vehicles) can perform post-quantum cryptography without excessive energy drain.\\

\item \textbf{Adaptive Security Policies and Protocols:} Future TCPS should employ adaptive security measures that can respond to changing conditions and threats. Instead of fixed cryptographic strength for security, we envision policies that adjust cryptographic strength and behaviors on the fly. For example, a vehicle might use faster but slightly less robust cryptography during normal operation for efficiency, but automatically switch to maximum security mode if an anomaly is detected or if it enters a high-risk location. Developing the framework for such adaptive security requires setting triggers and thresholds and ensuring seamless switching without interrupting operations. This also includes cryptographic agility in the face of quantum advancements – policies should mandate periodic review of which algorithms are considered safe. Research into formal methods for verifying the stability and consistency of adaptive security so that adaptation itself doesn’t introduce vulnerabilities will also be important.\\

\item \textbf{AI-Enhanced Hybrid Cryptosystems:} As PQC is rolled out, it will coexist with classical crypto for some time to maintain compatibility and as a fail-safe. Hybrid cryptosystems are already being standardized for transitional use. AI can enhance these systems by intelligently managing the complexity – deciding when both schemes are needed versus when one can suffice, optimizing the combination to reduce overhead, and maybe even learning which cryptographic primitive is more trustworthy in a given scenario. One promising concept is an AI-based decision engine that learns trust levels for different algorithms: for instance, if it observes any irregularities or suspected weaknesses in one algorithm, it could favor the other algorithm in the hybrid pair. Eventually, when legacy algorithms are phased out, the same AI could assist in seamlessly turning off the old algorithms system-wide. Moreover, AI could enable multi-layered cryptographic defenses – where each critical message being protected by multiple layers of cryptographic primitives, where the AI can adapt the number of layers based on risk. This kind of intelligent layering could mitigate even unforeseen vulnerabilities, as it’s unlikely an adversary can simultaneously break several strong algorithms unless they have a full-scale quantum computer.\\

\item \textbf{Side-Channel and Attack Resilience through AI:}  Future work should also explore the use of AI to defend against implementation-level attacks on PQC. As PQC algorithms are deployed, attackers will likely attempt side-channel attacks. These include methods such as power analysis and timing attacks. Fault injection attacks may also be used. Some research has already demonstrated these types of attacks on lattice-based schemes \cite{Ravi2023}. AI can be utilized to detect such attacks in real-time as part of anomaly detection and potentially to adapt algorithms to be more side-channel resistant. Additionally, AI might discover new side-channel patterns by analyzing execution traces with machine learning, thereby informing developers how to better harden the implementations.\\ 
% The interplay of AI and cryptography in this defensive context is an open research frontier – success here would mean PQC implementations in vehicles that can “self-heal” or self-adjust when under attack, providing robust operational security.

\item \textbf{Field Testing and Pilot Deployments:} Finally, moving from theory to practice will require extensive testing of PQC and AI systems in real transportation environments. Future work should include pilot deployments of PQC-enabled V2X networks and AI-based security management in those pilots to gather data. Projects like updating a fleet of connected vehicles with hybrid certificates or equipping a highway corridor with post-quantum secure roadside units, will yield invaluable insights. These deployments will help refine algorithms and help uncovering new research challenges, validate AI models (e.g., ensuring low false positives in anomaly detection), and generally pave the way for standardizing quantum-safe TCPS communication protocols. Collaboration between the automotive industry, government transportation agencies, and the cybersecurity community will be vital to turn the envisioned AI-driven PQC solutions into deployed technology. Early successes in this direction will build confidence that the looming quantum threat can be managed without compromising the tremendous benefits that TCPS promise for society.\\

\end{enumerate}

\section{Conclusion}
\label{sec:7}

% In conclusion, the convergence of post-quantum cryptography and artificial intelligence offers a compelling path forward to secure the connected transportation systems of the future. By proactively embracing PQC algorithms and augmenting them with intelligent, adaptive security mechanisms, we can ensure that the smart vehicles and smart roads of tomorrow remain trustworthy and resilient against even the most advanced cyber adversaries. The road ahead includes technical challenges and research questions, but the destination – a quantum-secure, AI-enhanced TCPS—is well worth striving for, as it underpins the safety and privacy of transportation in the 21st century and beyond.

The findings of this chapter confirm that today’s TCPS attack surface is both wide and evolving. Safety-critical data flows—V2V beacons, SPaT broadcasts, over-the-air updates—are vulnerable to eavesdropping, spoofing, message forgery, and denial-of-service when adversaries can exploit weaknesses in classical cryptography or underlying network protocols. Shor’s and Grover’s algorithms raise this risk dramatically by rendering RSA, ECDSA, and other secure cryptographic schemes either breakable or substantially weakened, undermining the trust anchors that protect certificates, session keys, and firmware images. Even with quantum-safe symmetric key sizes, side-channel and replay attacks threaten message integrity and privacy, while RF jamming or protocol flooding can degrade channel availability. Thus, cyber-resilience in TCPS hinges on replacing vulnerable primitives, hardening implementations against data leakage, and deploying layered defences that maintain confidentiality, integrity, authenticity, availability, and privacy under both classical and quantum threat models. In short, without a comprehensive cryptographic upgrade, the very communications that make connected mobility safer could become vectors for catastrophic failure.\\

Post-quantum cryptography supplies that upgrade path by substituting hardness assumptions that are believed to resist quantum computers. Lattice-based KEMs, such as CRYSTALS-Kyber, lattice signatures, such as Dilithium and Falcon, hash-based SPHINCS+, and the code-based HQC scheme now form NIST’s first generation of standards. They preserve public-key functionality—key establishment and digital signatures—yet differ in performance, key size, and message overhead, demanding careful profiling before deployment. Kyber and Dilithium deliver the most balanced trade-off for bandwidth-constrained V2X messaging, whereas SPHINCS+ and HQC offer conservative security at the cost of larger artifacts. Early trials on traffic-signal controllers and on-board units show that, after parameter tuning, these schemes can satisfy the 100 ms safety-latency target. In worst-case situations, such as very busy intersections or dense platoons, the extra computation can overload the embedded CPUs, and the increased packet volume can congest the 5.9 GHz radio channel. Most importantly, PQC adoption must be staged and governed so that legacy algorithms are revoked only after quantum-safe roots of trust are fully operational. Thus, a multi-year road map is needed for the transportation sector to achieve this goal.\\

% AI techniques widen that roadmap by enabling cryptographic agility and proactive defense. 
By enhancing cryptographic agility and enabling proactive defense, AI can greatly assist the transition to PQC schemes. Reinforcement-learning agents can sense channel congestion, processor load, and threat posture, then switch among PQC parameter sets, compress keys, or offload lattice computations to roadside edge computers without human intervention. Currently, supervised models can flag anomalous V2X traffic patterns, side-channel leakage traces, or certificate misuse with millisecond response times, providing an automated first responder against implementation-level exploits. On the other hand, generative AI could augment protocol design by synthesizing realistic traffic scenarios, which can help engineers to conduct stress tests on new PQC stacks before field rollout. Forecasting models have the ability to correlate quantum-computing progress indicators with cryptanalytic risk so that operators can harden keys or rotate algorithms before attackers gain an advantage. Overall, these AI layers bridge the gap between static cryptographic policies and the dynamic need of real-world mobility, transforming security into a continuously optimized control loop. Thus, AI-enabled PQC transforms TCPS from a reactive posture to a self-adapting, intelligence-driven security architecture.\\

Looking forward, research should converge on several priorities. First, lightweight PQC primitives and hardware accelerators are necessary to meet the low-latency budget of safety-critical wireless communication on resource-constrained OBUs and sensors. Second, formally verified adaptive-security policies should facilitate seamless fallback and upgrade paths as algorithms evolve and quantum timelines become more precise. Third, hybrid cryptosystems require rigorous deployment playbooks: determining when to transmit dual signatures, sequencing trust-anchor retirement, and auditing residual classical keys. Fourth, large-scale pilot corridors should integrate PQC stacks, AI controllers, and certificate infrastructures under actual roadway traffic loads, supplying empirical data that can refine standards. Cross-disciplinary collaboration—among cryptographers, AI researchers, transportation engineers, and regulators—will be the key to this agenda. If these research fronts advance in parallel, the vision of a quantum-safe and AI-hardened TCPS that could deliver safe, efficient, and secure mobility can be realized well before large-scale quantum adversaries emerge.
%%%%%%%%%%%%%%%%%%%%%%%% referenc.tex %%%%%%%%%%%%%%%%%%%%%%%%%%%%%%
% sample references
% %
% Use this file as a template for your own input.
%
%%%%%%%%%%%%%%%%%%%%%%%% Springer-Verlag %%%%%%%%%%%%%%%%%%%%%%%%%%
%
% BibTeX users please use
% \bibliographystyle{}
% \bibliography{}
%
%\biblstarthook{

% -------------------------------------------------

\end{document}